\documentclass[twocolumn,english,prl]{revtex4-1}
\usepackage{color}
\usepackage{graphicx}
\usepackage[T1]{fontenc}
\usepackage{float}
\usepackage{textcomp}
\usepackage{amsmath}
\usepackage{amssymb}
\usepackage{graphicx}
\usepackage{esint}
\usepackage{natbib}
\usepackage{color}

\begin{document}


\title{Persistent currents in topological and trivial confinement defined in silicene}


\author{Piotr Jurkowski}

\affiliation{AGH University of Science and Technology, Faculty of Physics and
Applied Computer Science,\\
 al. Mickiewicza 30, 30-059 Krak\'ow, Poland}

\author{Bart\l{}omiej Szafran}

\affiliation{AGH University of Science and Technology, Faculty of Physics and
Applied Computer Science,\\
 al. Mickiewicza 30, 30-059 Krak\'ow, Poland}

\begin{abstract}
We consider states bound at the flip of the electric field in buckled silicene. Along the electric flip lines a topological confinement is formed with the orientation of the charge current and the resulting magnetic dipole moment determined by the valley index. We compare the topological confinement to the trivial one that is due to a local reduction of the vertical electric field but without energy gap inversion. For the latter the valley does not protect the orientation of the magnetic dipole moment from inversion by external magnetic field. We demonstrate that the topologically confined states can couple and form extended bonding or antibonding orbitals with  the energy splitting influenced by the geometry and the external magnetic field.
\end{abstract}

\maketitle

\section{introduction}

A clean electrostatic confinement of charge carriers in quantum dots  provides environment
for precise studies of localized states,  energy spectra, coherence times \cite{eqd} electron-electron interactions,  \cite{reim} as well as for manipulation of the charge \cite{ceqd1,ceqd2}, spin \cite{eqds} and valley \cite{blgqd4} degrees of freedom.
In gapless graphene the carrier confinement by electrostatic potentials is excluded 
by the Klein tunneling \cite{kte}. 
The electrostatic confinement becomes available when the energy gap is open by vertical electric field: in bilayer graphene \cite{blgp1,blgp2,blgp3,blgp4,blgp5} and in silicene  \cite{ni,Drummond12}.
Silicene is an atomic monolayer graphene-like material \cite{Molle17,rev1,chow,Liu11,Ezawa}  
with buckled \cite{Molle17} crystal lattice. 
The spatial control of the energy gap by gating allows for electrostatic confinement of charge carriers and formation of quantum-dot bound states with discrete energy spectra 
-- see Refs. \cite{blgqd,blgqd1,blgqd2,blgqd3,blgqd4} for bilayer graphene and Ref. \cite{scirep} for silicene. 
 
In both bilayer graphene and in silicene the energy gap can be locally inverted by the flip of the electric field vector. The flip of the vertical electric field  forms a topological confinement of chiral currents along the zero line of the 
symmetry breaking potential \cite{morpugo,macdo,Ezawa12a,kink}
with bands that appear within the energy gap. For bilayer graphene this confinement is also achieved at the stacking domain walls  induced by line defect \cite{down,prx} or twist of the layers \cite{twi,margi}. 
In silicene \cite{Ezawa12a} and staggered monolayer graphene \cite{kink} the topological band is single and linear as a function of the wave vector  while two non-linear bands appear in  bilayer graphene \cite{morpugo,macdo,down,kink}. The reflectionless one-dimensional channels that appear with the flip of the electric field \cite{morpugo,macdo,Ezawa12a,szufran,kink} are similar to
the edge channels in the quantum Hall spin insulators \cite{Hasan10,Badarson13,Qi11} only with the valley degree of freedom replacing the spin in protection  mechanism against backscattering.
Similar confinement of unidirectional currents in the bulk of the monolayer graphene is observed at n-p junctions but only at strong magnetic fields \cite{np1,np2,np3,np4,np5,np6,np7}.

Here, we consider quasi zero-dimensional states localized along closed lines of the flip of the vertical electric field in buckled silicene.
The states appear within a locally vanishing energy gap.
We find that the chiral nature of the confined states is revealed by the direction of  current  circulation around the zero lines that is strictly related to the valley degree of freedom of the confined states. When the external magnetic field is applied the sign of the energy response depends only on the valley state.
Similarly, the current in the topological confinement cannot be reoriented for a given state by the external magnetic field, unlike the persistent currents \cite{pc2,pc5} for metal  
\cite{pc1,pc4}, semiconductor \cite{pc3,add1,add2} or etched graphene \cite{trin1,trin2,trin3} quantum rings.
 We compare the results for the topological confinement with the trivial one
resulting from the spatial variation of the energy gap without the inversion of the conductance and valence bands.  For the trivial confinement the external magnetic field can reorient the current. In this respect the loops of current at the trivial electrostatic confinement 
are similar to the ones flowing in etched graphene quantum rings \cite{trin1,trin2,trin3,trin4}.  
 We show that the topological confinement loops at  separate zero lines form   extended orbitals as in double quantum dots \cite{dqd}.

\section{Theory}

We consider a buckled silicene monolayer in inhomogenous electric field.
We consider first the system with a circular symmetry (see Fig. 1) with the potential bias between the sublattices that changes along the radial direction. We set the potential  
at the $A$ sublattice 
\begin{equation}
V_A({\bf r})=V_g \left(1-2\exp(-r^4/R^4)\right), \label{po2}
\end{equation}
and assume that for the silicene  placed symmetrically between the gates the potential on the $B$ sublattice is opposite $V_B({\bf r})=-V_A({\bf r})$. 
For potential given by Eq. (\ref{po2}) the electric field changes orientation at a distance $r=\ln(2)^{1/4}R$ from the origin.
For the negative potential on the $A$ sublattice in the potential center the $K$ ($K'$) the electron currents flow clockwise (counterclockwise) along the flip of the electric field \cite{szufran}. 

\begin{figure}
\begin{tabular}{ll}
\includegraphics[width=0.5\columnwidth]{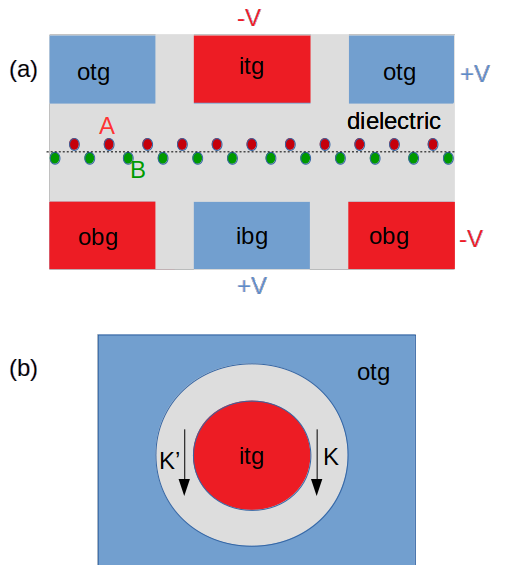} &
 \includegraphics[width=0.35\columnwidth]{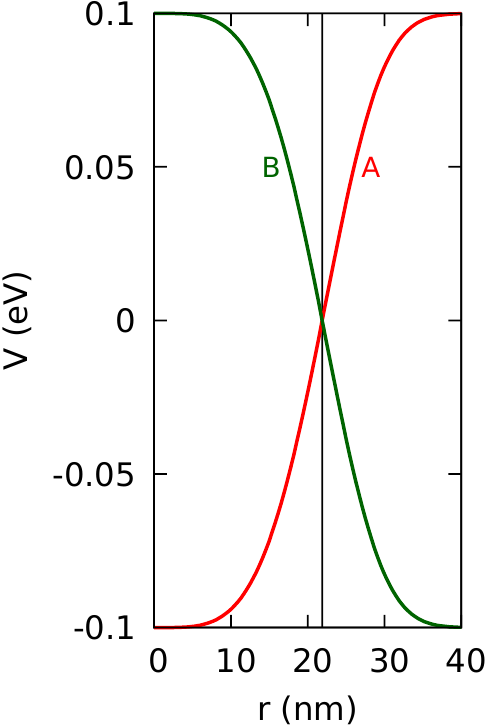} \put(-75,105){(c)}  
\end{tabular}
\caption{(a) Side view of the considered system. A buckled silicene layer (green and red dots stand for the Si atoms at the $B$ and $A$ sublattice respectively) is embedded
within a dielectric (grey area) in a system of metal gates inducing negative (red) and positive (blue) potential energy for electrons. The sublattice $A$ ($B$) is closer to the top gates (bottom gates). 'obg' and 'ibg' ('otg' and 'itg') stand for outer and inner bottom (top) gates. 
Panel (b) shows the top view of the system from the $A$ sublattice side. 
In (c) we plot the model potential on $A$ (red) and $B$ (green) sublattices for potential
of a rotational symmetry [Eq.(1)] with $r$ standing for the distance from the origin, for $V_g=0.1$ eV and $R=24$ nm. The potential on the $B$ sublattice is opposite $V_B=-V_A$.
The arrows in (b) show the direction of the valley polarized electron currents for states confined at the electric field flip. The $K'$ ($K$) electrons move with the negative potential on the $A$ sublattice on the left (right) hand side, i.e. counterclockwise (clockwise). 
}\label{scch}
\end{figure}
\begin{figure}
\begin{tabular}{l}
\includegraphics[width=0.9\columnwidth]{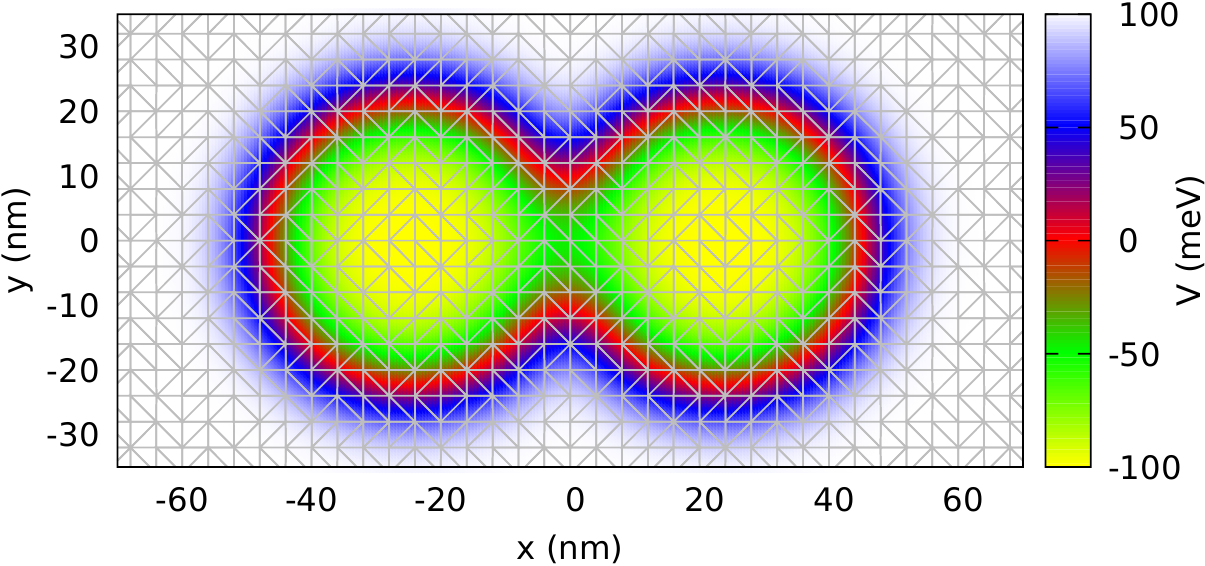}  \end{tabular}
\caption{The electrostatic potential on the $A$ sublattice and a mesh of the triangular elements for the finite element method (right square equilateral triangles of leg length 5 nm) with potential given by Eq. (\ref{po2}) with $l=R=24$ nm. 
The vertical electric field vanishes within the red area. 
 A fragment of the computational box is displayed.
In our implementation of the finite element method the solution in each triangle is spanned in the basis of 6 Lagrange shape functions with the nodes on the corners of the triangle and in the center of each side.}\label{siatka}
\end{figure}

\subsection{Atomistic tight-binding Hamiltonian}

The states in both circular and lower symmetry potentials are also analyzed with the atomistic tight-binding Hamiltonian \cite{Liu11,Ezawa,chow}, 
\begin{eqnarray}
H_{TB}&=&-t\sum_{\langle k,l\rangle } p_{kl} c_{k}^\dagger c_{l} 
+it_{SO} \sigma_z\sum_{\langle \langle k,l\rangle \rangle  } p_{kl} \nu_{kl} c^\dagger_{k} c_{l} \nonumber 
\\ && +\sum_{k} V_k c^\dagger_{k}c_{k}+\frac{g\mu_B B}{2}\sigma_z. \label{hb0}
\end{eqnarray}
The first sum describes the nearest neighbor hopping. The second sum is the atomistic form of the intrinsic
spin-orbit interaction \cite{km} with $\nu_{kl}=\pm 1$. The sign of $\nu_{kl}$  is  positive (negative)  for  the next nearest neighbor hopping
 via the common neighbor ion that turns
 counterclockwise (clockwise) and  $p_{kl}$ is the Peierls phase that introduces the magnetic field
$p_{kl}=e^{i\frac{e}{\hbar}\int_{\vec{r_k}}^{\vec{r_l}}\vec {\bf A}\cdot \vec {dl}}$,
where ${\bf A}$ is the vector potential.
We use the  symmetric gauge ${\bf A}=(-By/2,Bx/2,0)$ for the magnetic field perpendicular to the silicene lattice $(0,0,B)$. 
The tight-binding nearest-neighbor hopping Hamiltonian is $t=1.6$ eV \cite{Liu11,Ezawa},
and  $t_{SO}=3.9$ meV is the intrinsic spin-orbit coupling constant \cite{Liu11,Ezawa}. 
The positions of the ions of the $A$ sublattice 
${\bf r}_{\bf k}^A=k_1 {\bf a}_1+k_2 {\bf a}_2$  are generated 
with the crystal lattice vectors
${\bf a}_1=a \left(\frac{1}{2},\frac{\sqrt{3}}{2},0\right)$
and ${\bf a}_2=a \left(1,0,0\right)$, where $a=3.89$ \AA\; is the silicene lattice constant, and $k_1$, $k_2$ are integers. 
The $B$ sublattice ions are generated by ${\bf r}_{\bf k}^B={\bf r}_{\bf k}^A+(0,d,\delta)$,
with the in-plane nearest neighbor distance
$d=2.25$ \AA, and the vertical shift of the sublattices $\delta=0.46$ \AA.
In Eq. (\ref{hb0}) $V_k$ is the electrostatic potential on ion $k$. The intrinsic spin-orbit coupling is diagonal in 
the basis of $\sigma_z$ eigenstates, therefore the $z$ component of the spin is 
used as a quantum number below.

The Hamiltonian (2) can be rewritten in a compact form 
\begin{equation} H_{TB}=\sum_{k,l} h_{kl} c_k^\dagger c_l, \end{equation} where the elements $h_{kl}$ are defined by Eq. (2). 
For the eigenfunction $\Psi$ of the atomistic Hamiltonian, with the value of $\Psi_l$ on ion $l$, the electron current flowing from ion $l$ to ion $k$, as derived \cite{der} from the Schr\"odinger equation is \begin{equation}
{\bf J}_{lj}=\frac{i}{\hbar} \left( h_{lj}\Psi_l^* \Psi_j - h_{jl} \Psi_l \Psi_j^*\right).\label{pro} \end{equation}
For Hamiltonian eigenstates Eq. (\ref{pro}) provides the probability current flow which is persistent as a characteristic property of a stationary state. 
The persistent charge current for a given stationary state has the opposite orientation to the probability  current. Since the intrinsic spin-orbit coupling is diagonal in $\sigma_z$ spin component, the considerd currents are spin-polarized in the perpendicular magnetic field. The
considered system does not contain short-range scatterers, so that the currents are also valley polarized, at least for magnetic fields for which the valley degeneracy is lifted.

\subsection{Continuum Hamiltonian}

The continuum Hamiltonian is used to determine the valley and angular momentum (when available) of the eigenstates calculated in the atomistic approach. The continuum Hamiltonian is a low-energy approximation to the atomistic Hamiltonian.
In the low-energy approximation the carriers are described by a spinor wave function with
components defined on $A$ and $B$ sublattices of the silicene crystal lattice
$\psi=\left( \begin{matrix} \psi_A & \psi_B\end{matrix}\right)^T$.
The low-energy approximation to the atomistic tight-binding Hamiltonian  \cite{Liu11} reads
\begin{eqnarray} 
H_\eta & =&  \hbar v_F \left( \begin{array}{cc} 0 & k_x+\eta i k_y \\ 
k_x-\eta i k_y &  0\end{array} \right) \nonumber\\
& + & \left( \begin{array}{cc} V_A(x,y)  + \eta  \sigma_z t_{SO} & 0 \\  \nonumber
0  & V_B(x,y)  -\eta  \sigma_z t_{SO} \end{array} \right)  \\ \label{hamu} 
&+& \frac{g\mu_B B}{2}\sigma_z I,
\end{eqnarray}
where $\eta$ stands for  the valley index ($\eta=1$ for the $K$ valley and $\eta=-1$  for the $K'$ valley), $I$ is the identity matrix,
 ${\bf k}=-i\nabla +\frac{e}{\hbar}{\bf A}$.
In Eq. (\ref{hamu})
$v_F=3dt/2\hbar$,
is the Fermi velocity.
\subsection{Circular potentials}

For circular potentials $V_A(x,y)=V_A(r)$ the Hamiltonian eigenfunctions 
can be labeled by an integer magnetic quantum number $m$,
 \begin{equation}
\Psi_{m,\eta}= \left(\begin{array}{c} f_A (r) \exp(im\phi) \\ f_B (r) \exp(\exp(i(m-\eta)\phi)\end{array} \right), \label{laal}
\end{equation} where $f_A(r)$ and $f_B(r)$ are the radial functions on the sublattices.
We take a circular flake of radius $R=60$ nm. At the edge of the flake
we  apply the zigzag boundary conditions \cite{sweep}.
In order to avoid the fermion doubling problem we use an asymmetric finite difference quotient for the first derivative $f'=\frac{f(r)-f(r-dr)}{dr}$ instead the symmetric one \cite{sweep}. 
The Hamiltonian eigenequation with this quotient can be transformed into a scheme
that derives $f_A(r-dr)$, and $f_B(r-dr)$  from $f_A(r)$ and $f_B(r)$,
\begin{widetext}
\begin{eqnarray}
      f_A(r-dr)&=&\frac{dr}{ i \hbar v_F} \left(E-V_B(r)+\eta t_{SO} \sigma_z -\frac{g\mu_B B}{2}\sigma_z\right)f_B(r)+\left(1+\frac{dr\eta m}{r}+\frac{eBr\eta dr}{2\hbar}\right)f_A(r),\\
      f_B(r-dr)&=&\frac{dr}{ i \hbar v_F} \left(E-V_A(r)-\eta t_{SO} \sigma_z -\frac{g\mu_B B}{2}\sigma_z\right)f_A(r)+\left(1-\frac{dr\eta (m-\eta)}{r}-\frac{eBr\eta dr}{2\hbar}\right)f_B(r).
 \label{chc2}
\end{eqnarray}
\end{widetext}
The energies of the bound states are determined by the asymptotic condition
to be fulfilled at the origin $r=0$, which requires that \cite{sweep}
$f_A$ and/or $f_B$ function vanish at the origin when $m\neq 0$ and/or $m-\eta \neq 0$, respectively.

\subsection{Non-circular potentials and a finite element method}

For coupled systems we perform calculations also for lower symmetry potentials. We take the potential on the $A$ sublattice in form
\begin{equation}
V_A^{(2)}({\bf r})=V_g \left[1-2 e^{-\frac{({\bf r}-(l,0,0))^4}{R^4}}-2 e^{-\frac{({\bf r}+(l,0,0))^4}{R^4}}\right], \label{ala}
\end{equation}
where $2l$ is the distance between the centers of  electric field inversion loops.
As above, the potential on the $B$ sublattice is taken opposite to the one at the $A$ sublattice. The potential profile 
on the $A$ sublattice is
displayed in Fig. \ref{siatka}. 
In order to evaluate the eigenstates in the continuum approach we use Hamiltonian (5) in Cartesian coordinates and
the finite element method with the triangular elements on both sublattices and the shape functions in form of the second degree Lagrange interpolation polynomials within each of the elements \cite{solin}. The 
elements are right-angle isosceles triangles with the leg length of 5 nm. The side length of a rectangular computational box is taken up to 180 nm. We work with up to 3528 elements.

In order to deal with the fermion doubling problem and remove the spurious states from the low-energy spectrum
 \cite{sweep} we introduce the 
Wilson term \cite{wi} to the Hamiltonian 
\begin{equation}
H_D=- W_D \nabla ^2 \tau_z,
\end{equation}
with the Wilson parameter $W_D=36$ meV nm$^2$. This value of the Wilson parameter removes the spurious states with a negligible influence on the actual smooth solutions of the Dirac equation.

\begin{figure*}
\begin{tabular}{lll}
(a) \includegraphics[width=0.55\columnwidth]{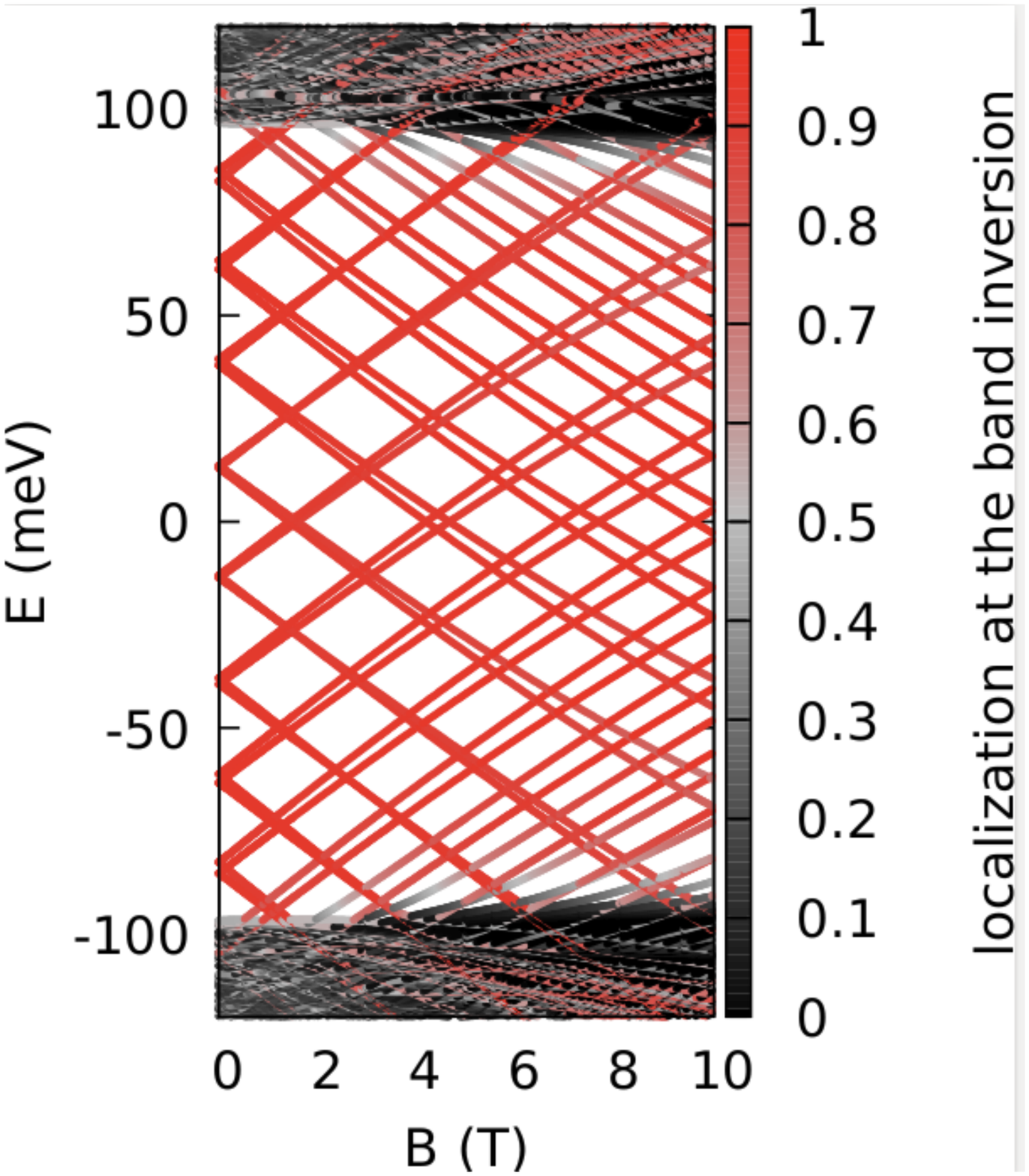} & (b) \includegraphics[width=0.44\columnwidth]{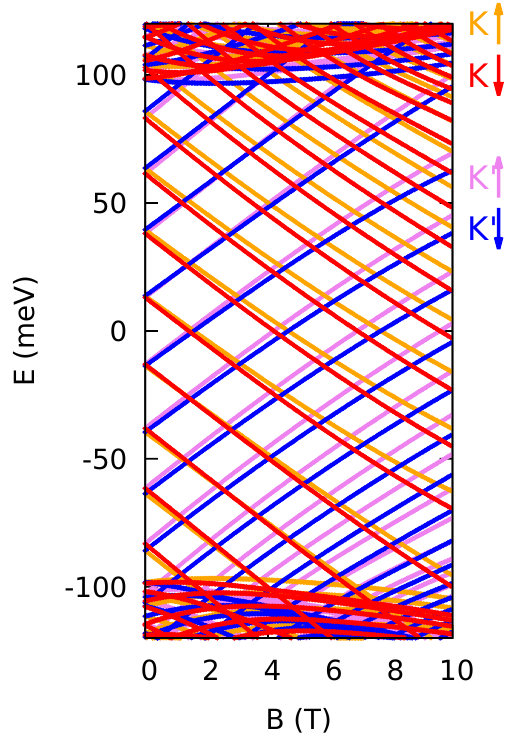} & (c) \includegraphics[width=0.44\columnwidth]{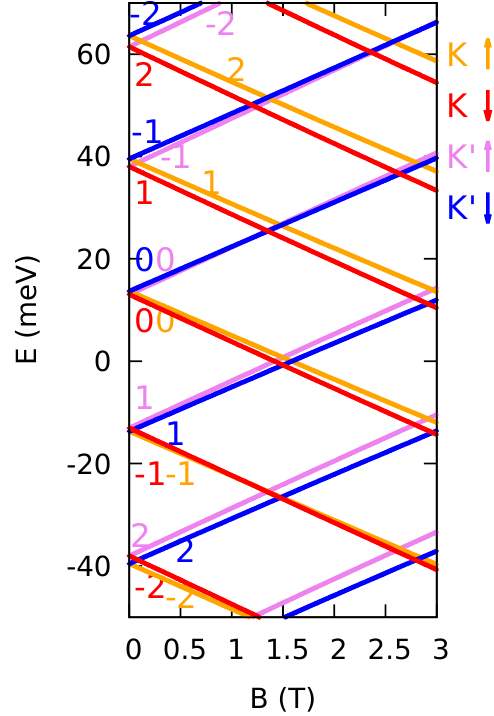} \end{tabular}
\begin{tabular}{lll}
(d) \includegraphics[width=0.75\columnwidth]{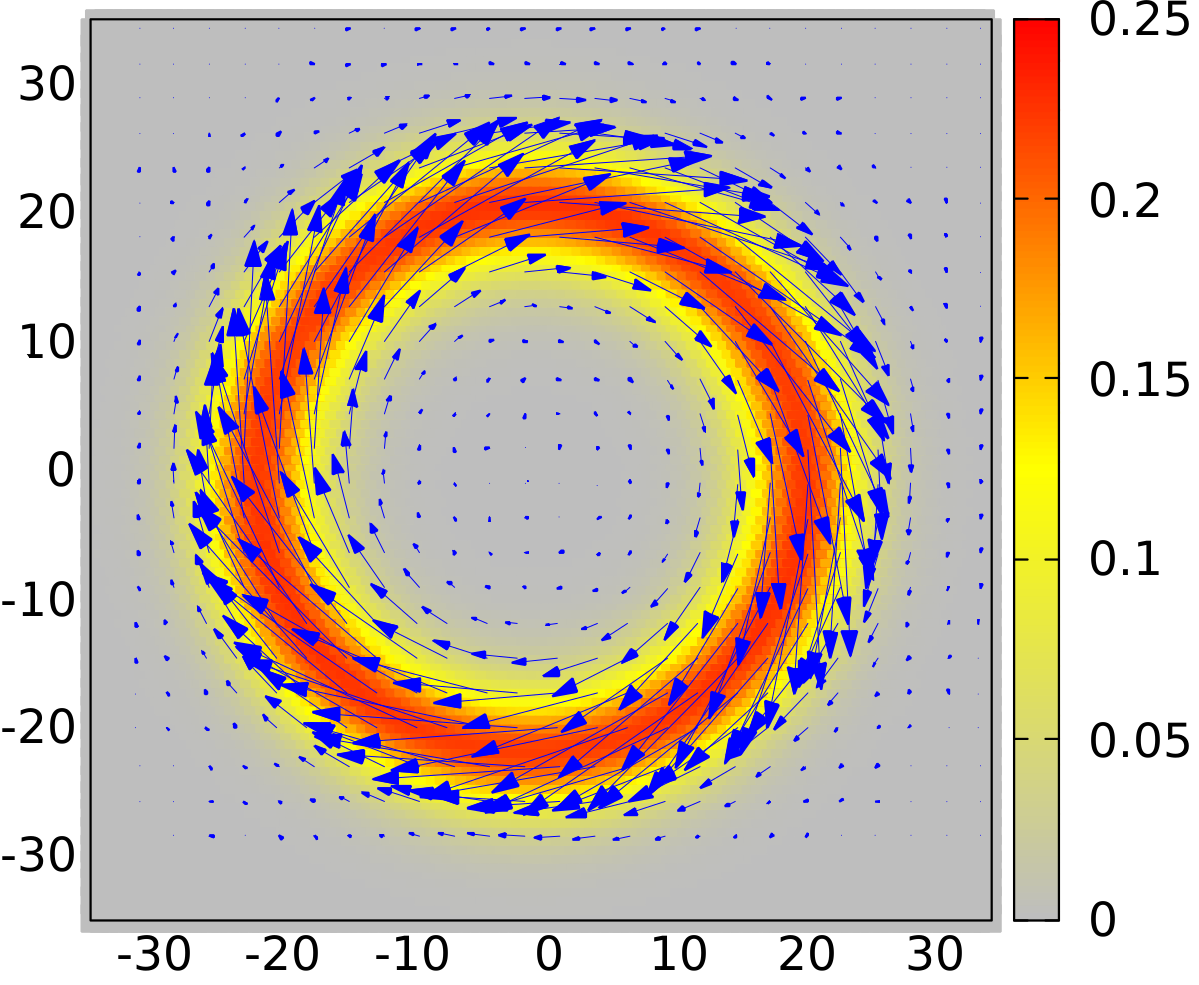}  \put(-200,60){\rotatebox{90}{y (nm)}} \put(-205,20){$K \downarrow$} \put(-115,-10){x (nm)} \put(-0,60){\rotatebox{90}{$|\Psi_A|^2$ (arb.un.)}} &\;\;\;\;\;\; & (e)  \includegraphics[width=0.75\columnwidth]{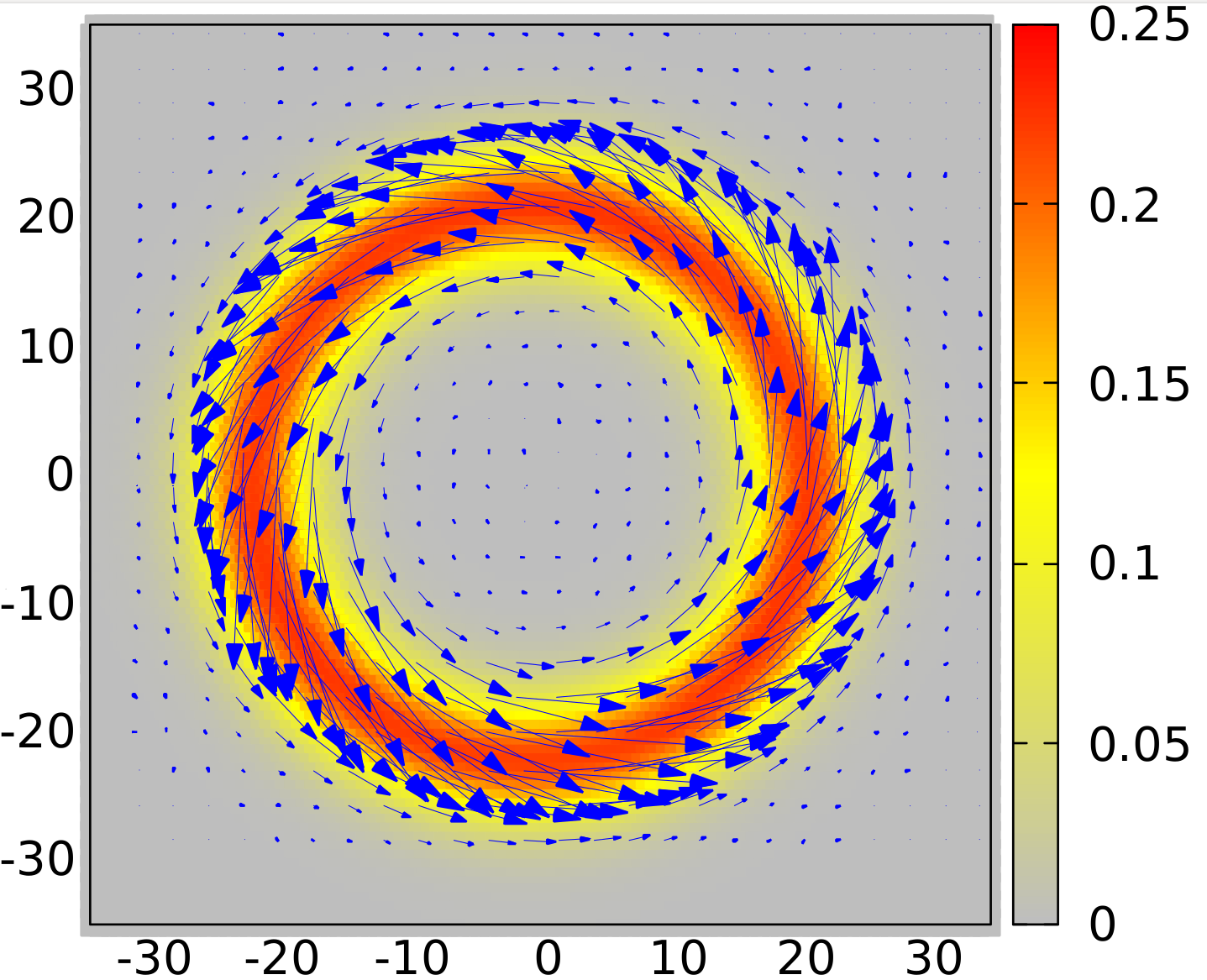}\put(-205,20){$K'
\downarrow$} \put(-115,-10){x (nm)} \put(-200,60){\rotatebox{90}{y (nm)}}\put(-0,60){\rotatebox{90}{$|\Psi_A|^2$ (arb.un.)}}
\end{tabular}
\caption{(a) The energy spectrum as calculated with the atomistic tight-binding approach. Color of the lines shows the localization of the states within the band inversion area calculated by integration of the probability density
within the annular area for $r\in(0.6R,1.4R)$. (b) The spectrum as calculated with the finite difference method.
Color of the lines indicates the spin and valley of the states.
The states with $|m|\leq 16$ are shown.
 (c) Zoom of (b) with the magnetic quantum numbers $m$ 
for the $A$ sublattice. (d) and (e) show the probability density and the probability current density for the first $K$ (d) and $K'$ (e) spin-down energy levels at $E>0$. } \label{circr}
\end{figure*}

\begin{figure}
\begin{tabular}{lll}
\includegraphics[width=0.29\columnwidth]{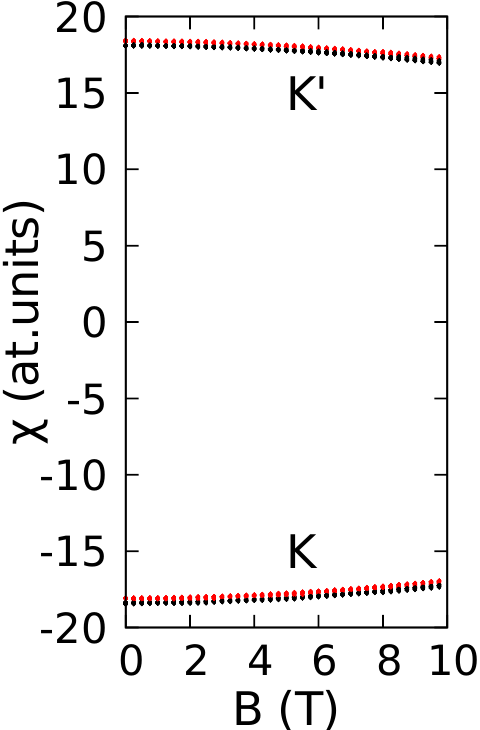} \put(-18,30){(a)} & \includegraphics[width=0.31\columnwidth]{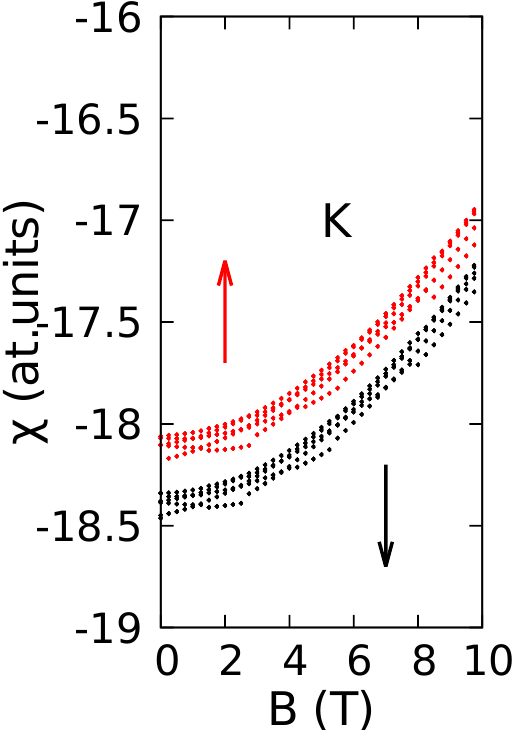} \put(-18,30){(b) } &
\includegraphics[width=0.31\columnwidth]{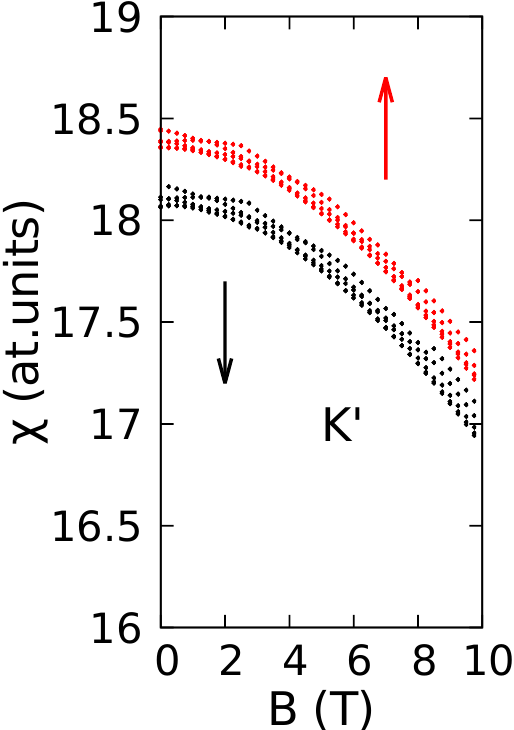} \put(-18,30){(c) }
 \end{tabular}
\caption{(a) The  current moment as given by Eq. (\ref{chi}) plotted for twenty states of Fig. 3(a)
with the smallest absolute values of the energy. The red (black) dots correspond
to the spin up (down) states. (b) and (c) show the zoom of the lower (higher) bands of panel (a).
} \label{chiju}
\end{figure}

\begin{figure}
\begin{tabular}{ll}
\includegraphics[width=0.475\columnwidth]{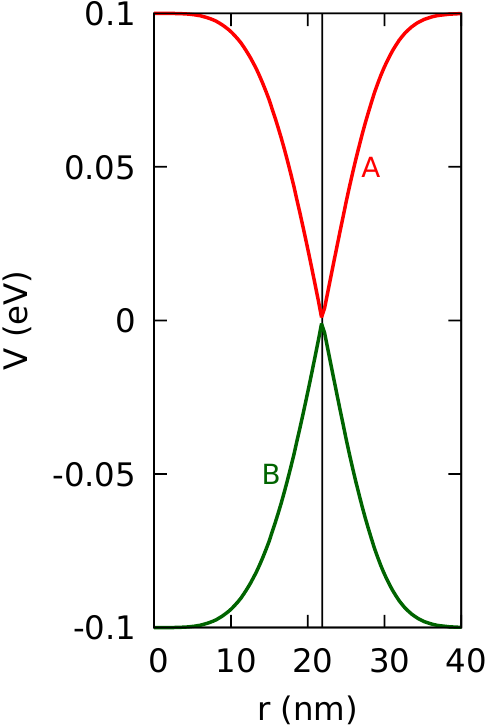}  \put(-20,35){(a)}  & \includegraphics[width=0.45\columnwidth,trim= 4cm 0 4cm 0 ,clip]{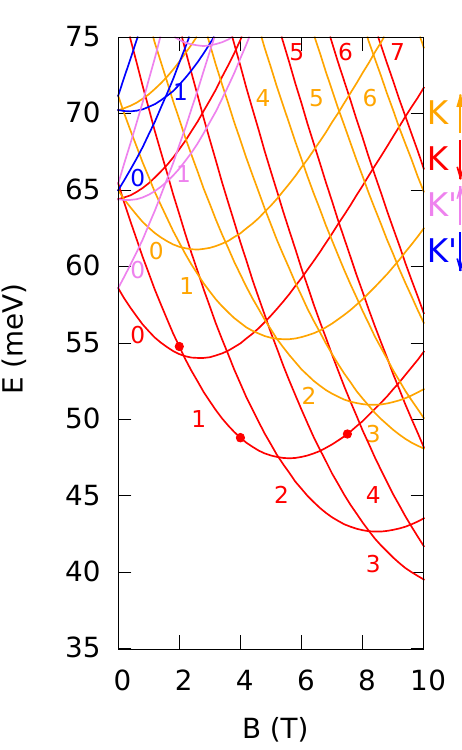} \put(-35,35){(b)}\\
\includegraphics[width=0.45\columnwidth]{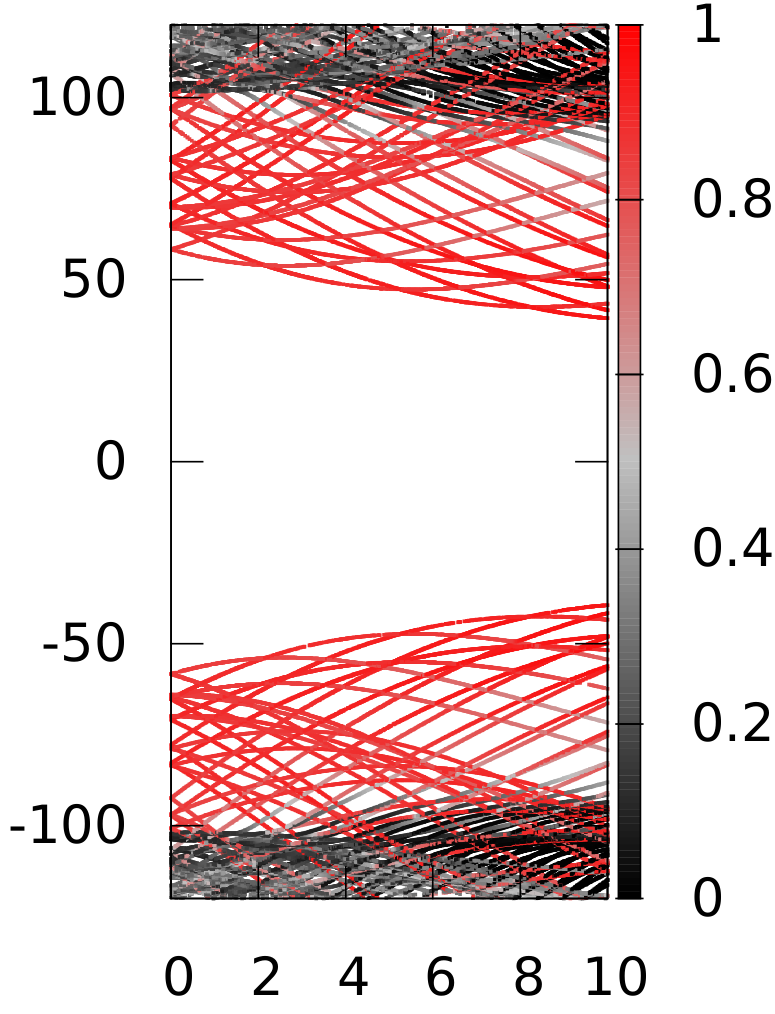}\put(-50,85){(c)} \put (-70,-8){B (T)} \put (-110,60){\rotatebox{90}{E (meV)}}&
\includegraphics[width=0.5\columnwidth,trim= 3.75cm 0.1cm 3cm 0 ,clip]{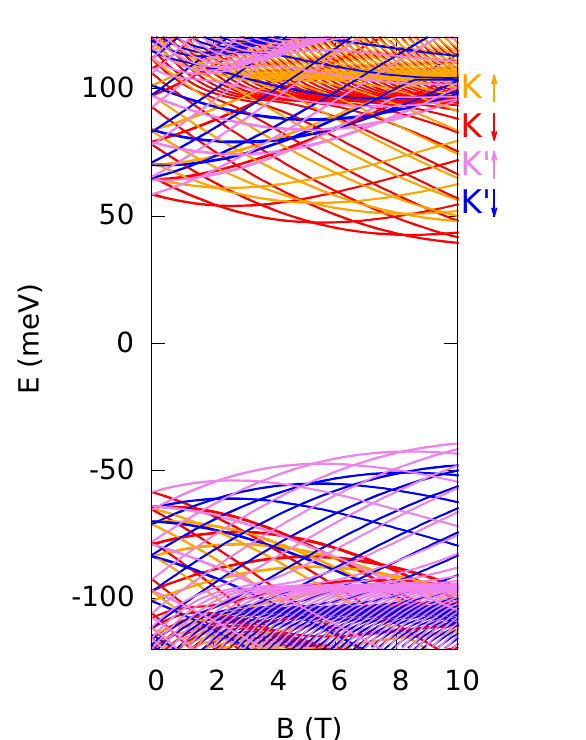}\put(-50,85){(d)} 

  \end{tabular}
\caption{(a) A trivial confinement potential given by Eq. (\ref{potr}) without the inversion of the energy gap (compare with Fig. 1(b) for the topological confinement). (b,d) The energy
spectrum calculated with the continuum model. Panel (b) is a zoom of Panel (d). 
(c) The results of the atomistic tight binding -- same as Fig. \ref{circr}(a) for the trivial potential. The color scale shows the integral of the probability density within the area $r\in(0.6R,1.4R)$.
 }\label{trivial}
\end{figure}

\begin{figure}
\begin{tabular}{lll}
\includegraphics[height=0.27\columnwidth]{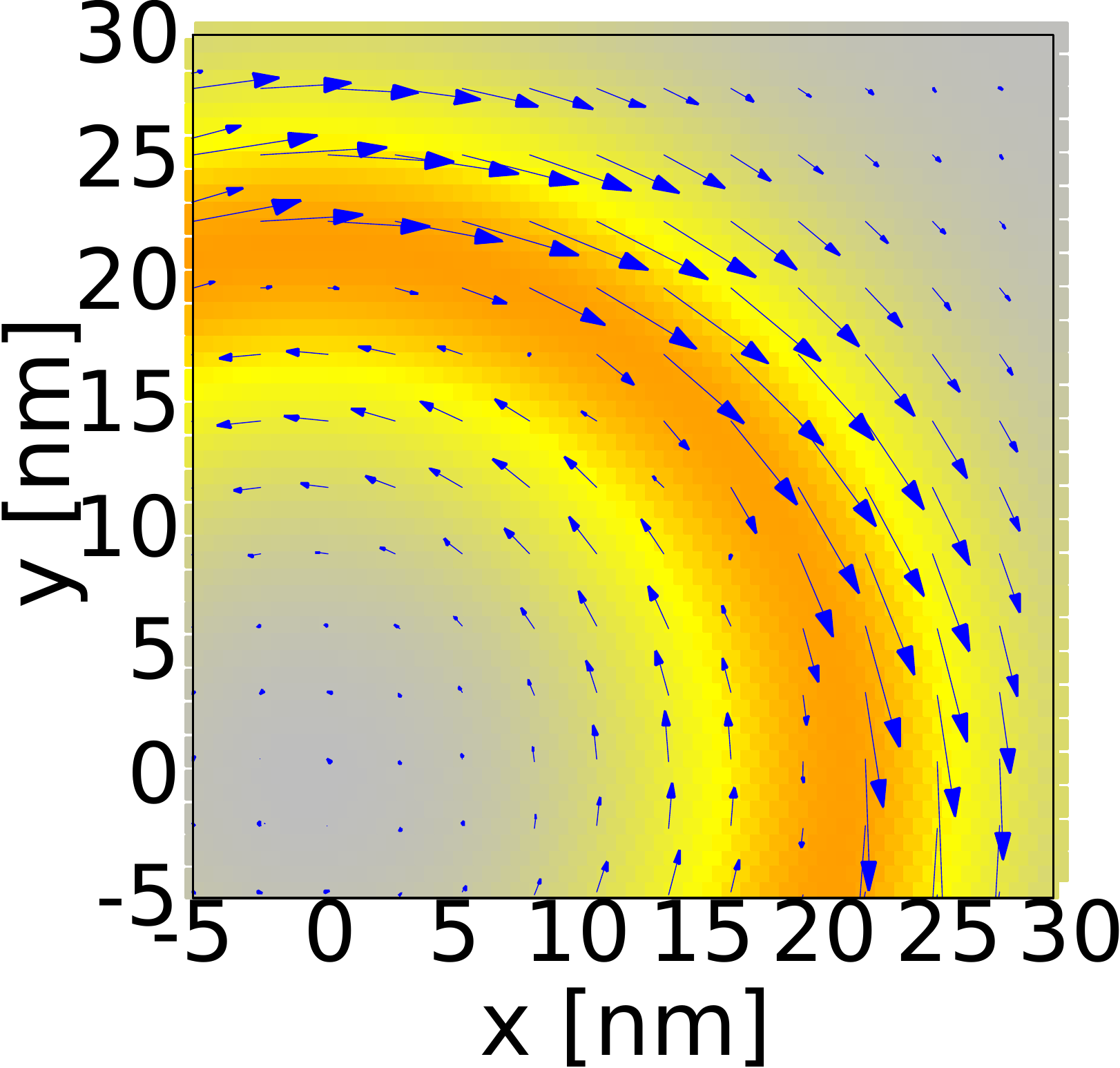} \put(-30,70){(a)} &
\includegraphics[height=0.27\columnwidth]{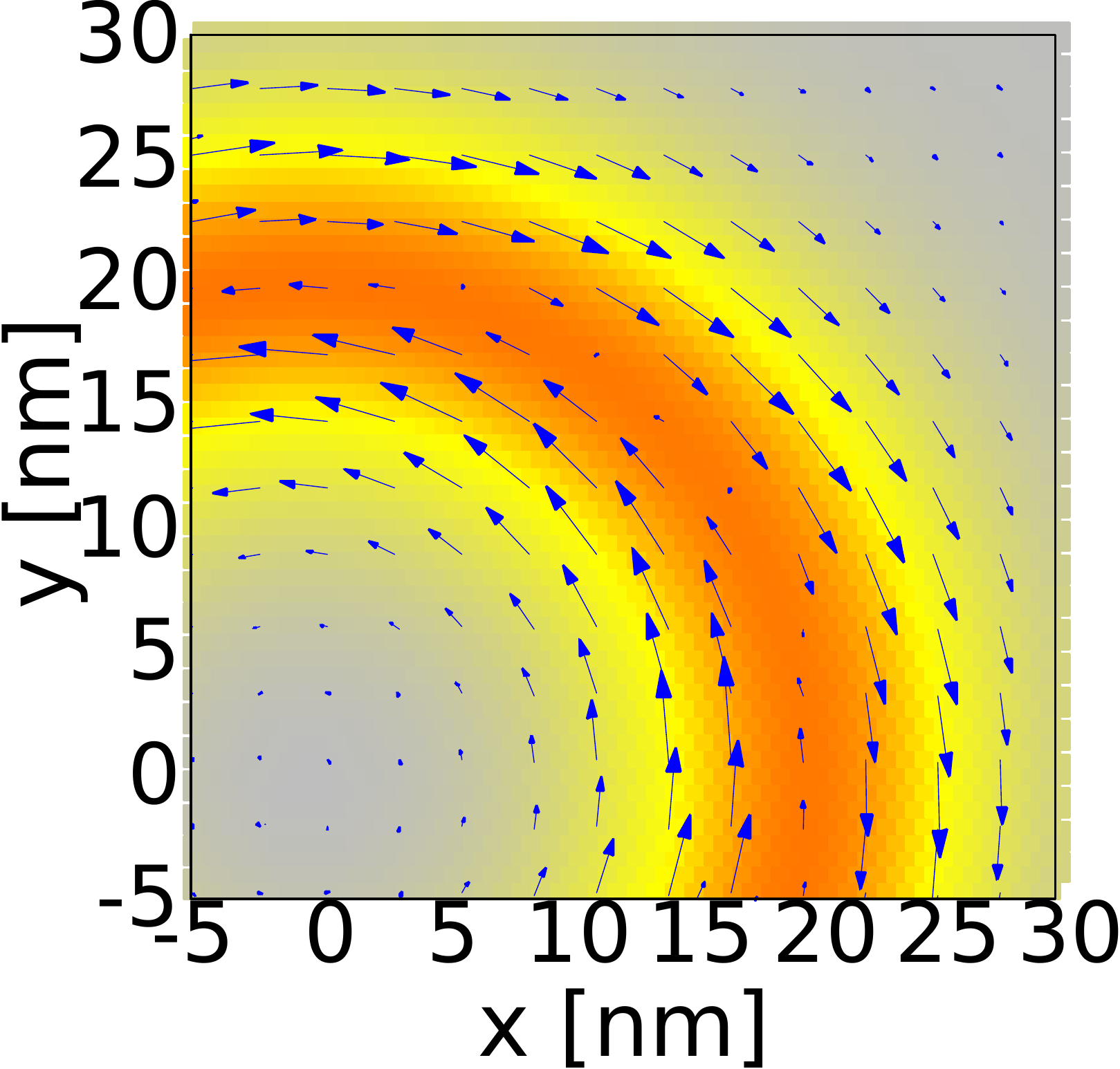}  \put(-30,70){(b)}&
\includegraphics[height=0.27\columnwidth]{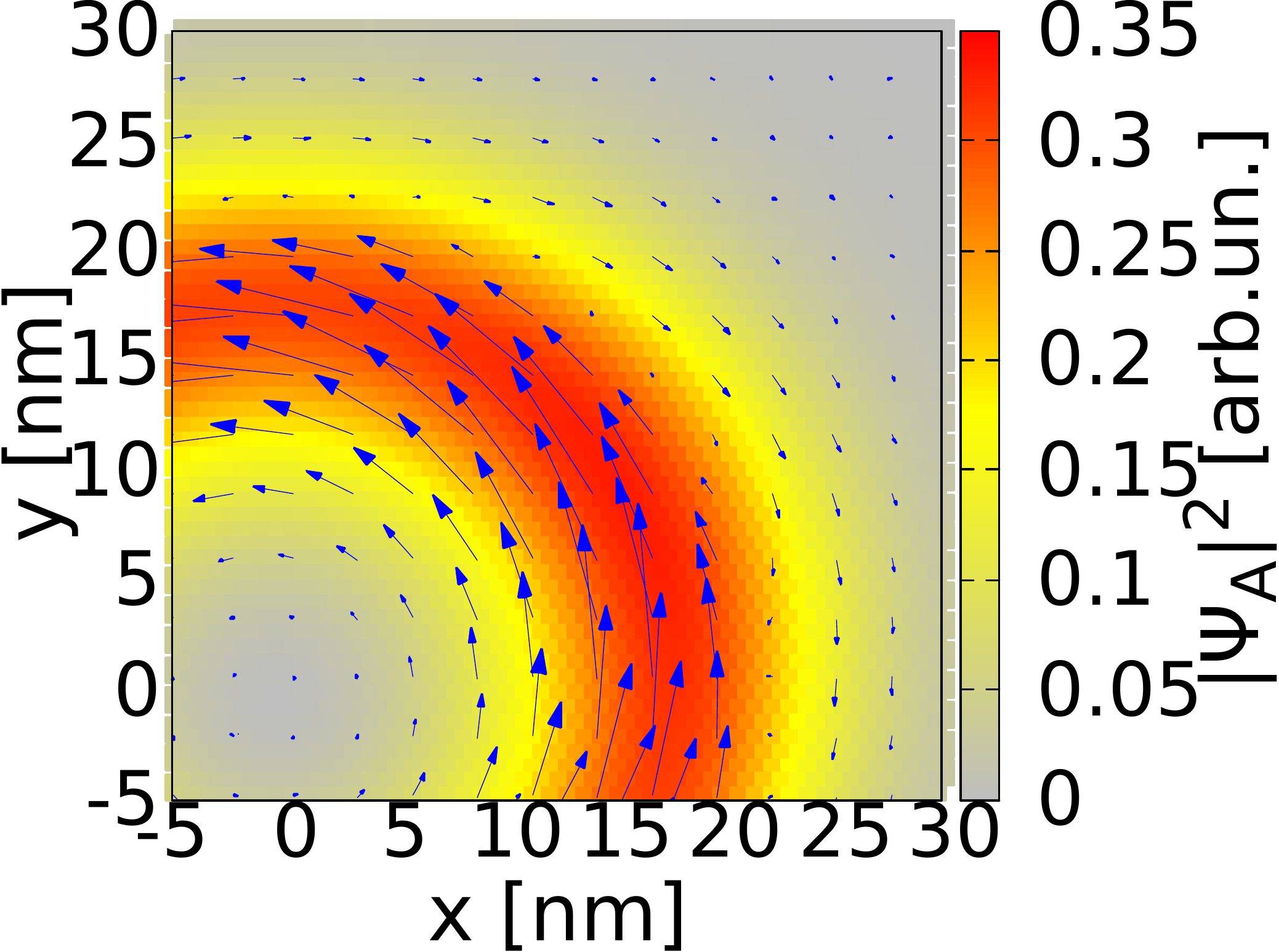}  \put(-30,70){(c)}\\    
  \end{tabular}
\caption{Electron density on the $A$ sublattice (color scale) and the electron current (vectors) for the $K\downarrow$ state with $m=1$ for $B=2$T (a), $B=4$T (b), $B=7.5$T (c) -- see the dots on this  energy level in Fig. \ref{trivial}(b) for the trivial confinement.
See Fig. \ref{chi9} for the current moment of $m=1$ state}\label{trivpra}
\end{figure}

\begin{figure}
\includegraphics[width=0.5\columnwidth]{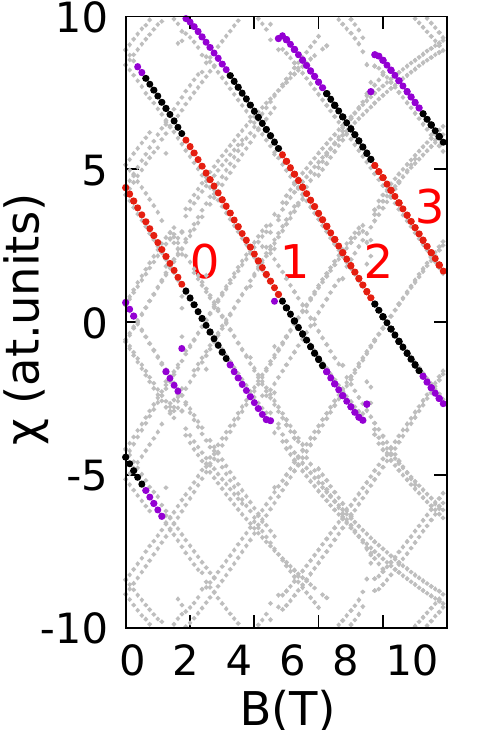} 
\caption{The current moment as given by Eq. (\ref{chi}) plotted for thirty states of Fig. \ref{trivial}(c) of the smallest absolute values of the energy. With the red, black, and purple dots we marked the lowest, second, and third lowest energy states on the $E>0$ side of the zero energy. By the integers we mark the $m$ values for the lowest-energy $K\downarrow$ states [cf. Fig. \ref{trivial}(b)].  
} \label{chi9}
\end{figure}

\begin{figure}
\begin{tabular}{ll}
\includegraphics[width=0.45\columnwidth]{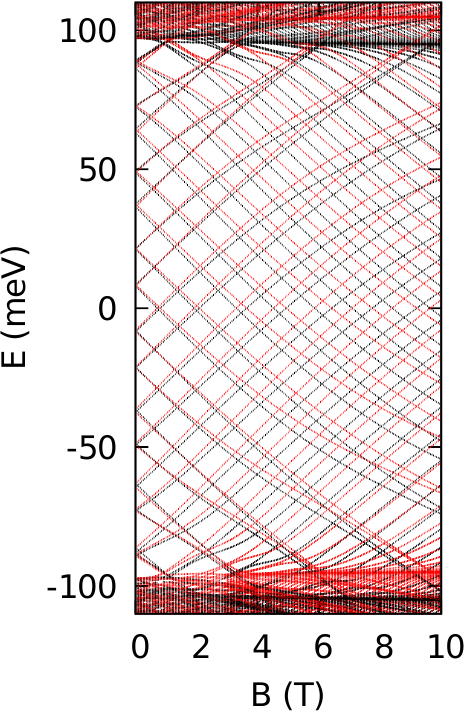} \put(-30,180){(a)} &\includegraphics[width=0.45\columnwidth]{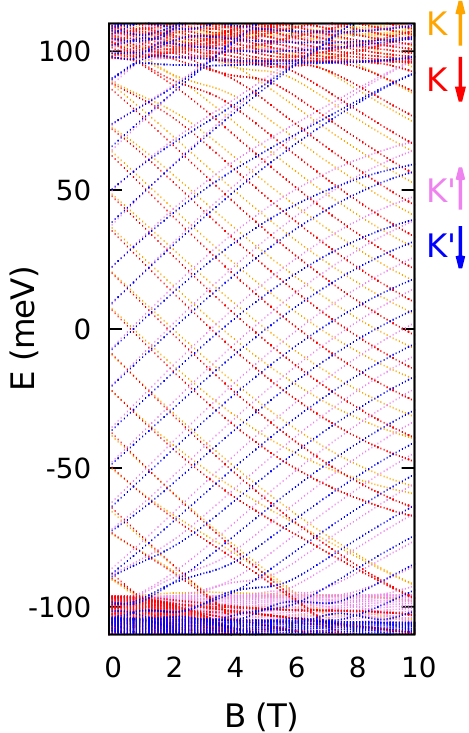}   \put(-30,180){(b)}  \\  \includegraphics[width=0.45\columnwidth]{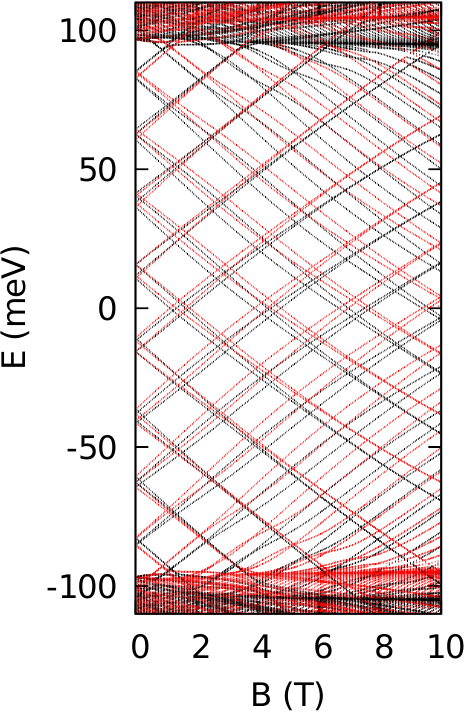}\put(-30,180){(c)}  &  \includegraphics[width=0.45\columnwidth]{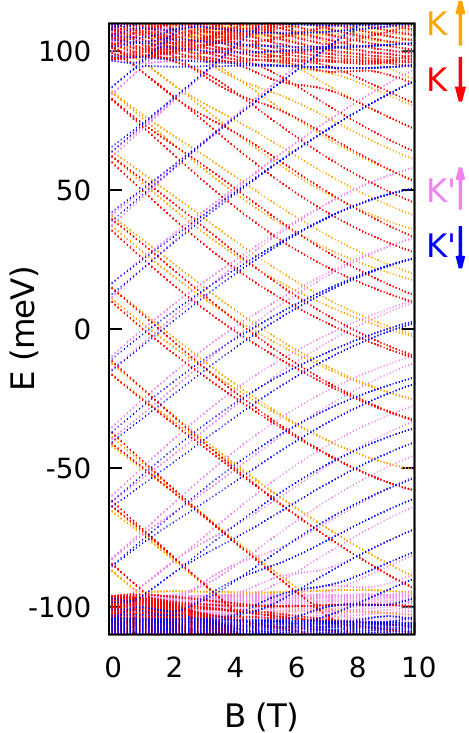}  \put(-30,180){(d)} 
\end{tabular}
\caption{The energy spectrum for the double-center system for potential given by Eq. (\ref{ala}) with $l=R$ (a,b) 
and $l=1.2R$ (c,d). The results of the atomistic tight binding are given in (a,c) with
the color standing for the spin (down -- black, up -- red). The finite-element solution to the continuum problem is
given in (b,d) with the color of the lines standing for the spin and valley of the energy level.  }
\label{xplain} \end{figure}

\begin{figure}
\includegraphics[width=0.6\columnwidth]{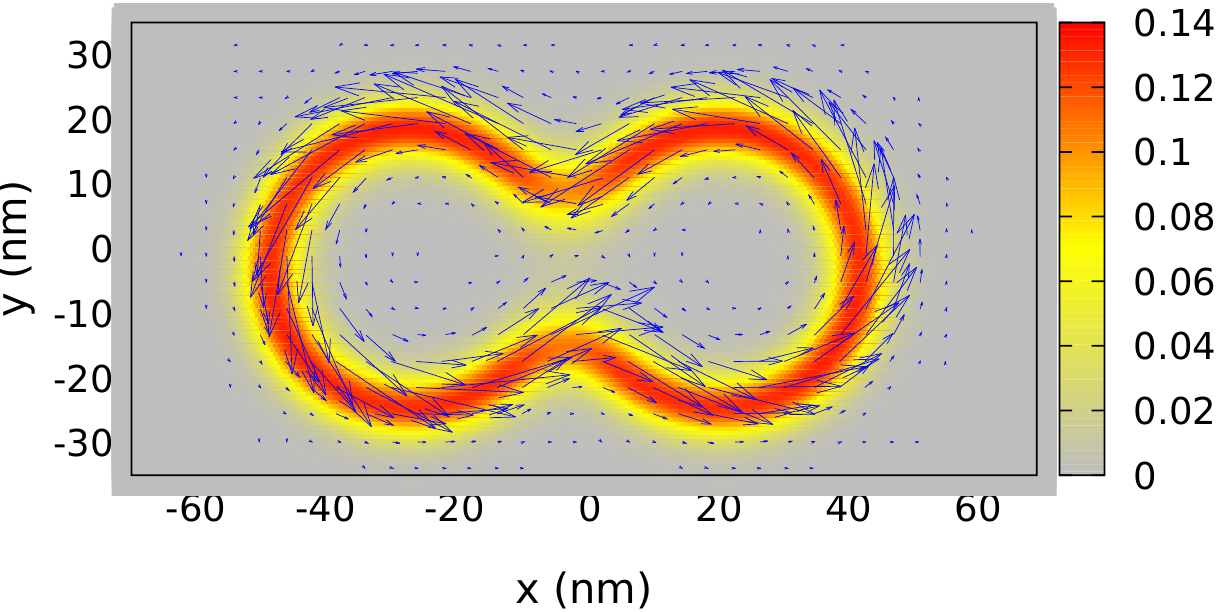}(a)
\includegraphics[width=0.6\columnwidth]{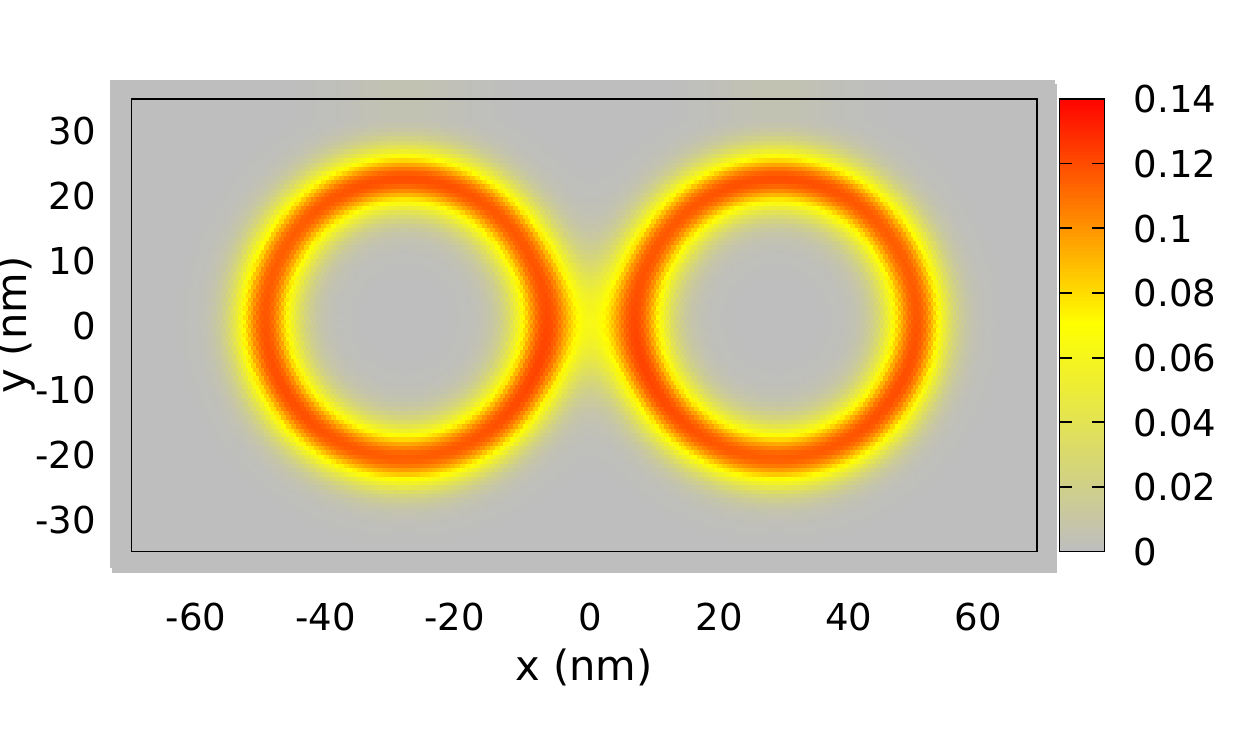}(b)
\includegraphics[width=0.6\columnwidth]{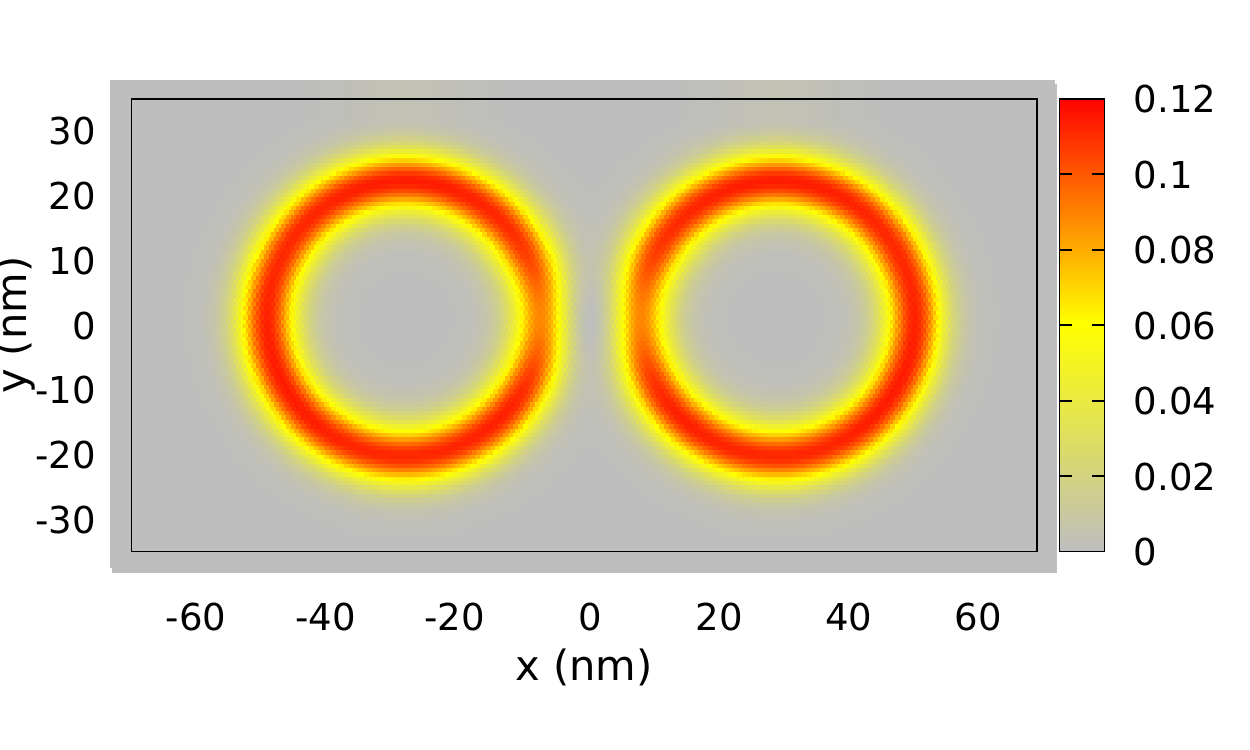}(c)    
\caption{(a) Same as Fig. \ref{circr}(e) but for the double ring system with $l=R$.
(b) The charge density at the $A$ sublattice for the lowest positive energy $K'\downarrow$ 
in the double ring system with $l=1.2$ R. (c) Same as (b) only for the next higher $K'\downarrow$  
energy level. The plots (a-c) were taken at $B=0$.} \label{next}
\end{figure}

\begin{figure}
\begin{tabular}{ll}
\includegraphics[width=0.45\columnwidth]{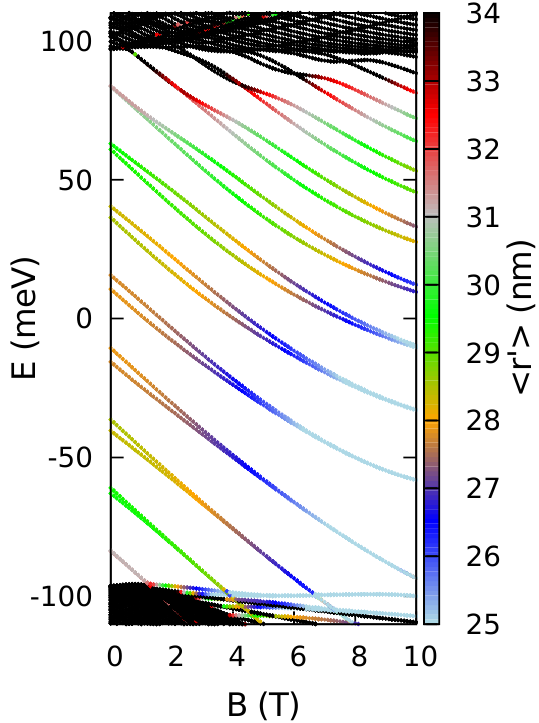} \put(-30,160){(a)} & \includegraphics[width=0.45\columnwidth]{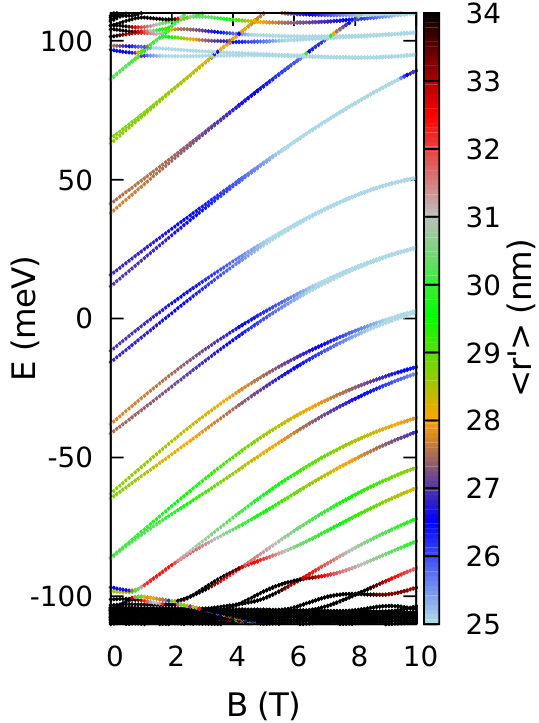} \put(-30,160){(b) }
 \end{tabular}
\caption{The energy levels of the $K$ (a) and $K'$ (b) energy levels 
for $l=R$ [taken of Fig. \ref{xplain}(b)] 
with the color of the lines indicating the average distance to the closer of the two centers 
of potential given by Eq. (\ref{ala}), $r'=\min(|{\bf r}-(l,0,0)|,|{\bf r}+(l,0,0)|)$. 
} \label{fju}
\end{figure}

\section{Results}

\subsection{Circular potential: topological confinement}

For the circular topological confinement we take the potential given by Eq. (1) and  take $V_g=0.1$ eV with  $R=24$ nm.  The results  are given in Fig. \ref{circr}.
Figure \ref{circr}(a) shows the results obtained with the atomistic tight binding
approach. The color of the lines corresponds to the localization of the wave function,
i.e. integral the probability density within the area $r\in(0.6R,1.4R)$.
Within the energy gap opened for $E\in(-V_g,V_g)$ we find a discrete energy spectrum.
All the states within the gap are localized near the zero line. 

The energy levels form quadruplets  at $B=0$, or more precisely, a pair of doublets split by the spin-orbit interaction of a few meV. The dependence of the energy levels on the magnetic field can be more easily explained
using the results of the continuum approach  of Fig. \ref{circr}(b). 
The energy levels of the discrete part of the spectrum agree very well with the results of the atomistic tight-binding approach [Fig. \ref{circr}(a)]. The continuum approach explicitly resolves the valley degree of freedom.
For the states that are not localized near the zero line, with the energy outside
the gap, the results differ, since the energy levels are localized either outside a
hexagonal silicene flake (atomistic tight-binding) or near the edges of a circular flake (continuum approach). 
The form of the boundary condition has no influence
on the confined states which are kept off the edge by the electrostatic potential \cite{scirep}.

In Fig. \ref{circr}(b) we can see that the $K'$ ($K$) energy levels increase (decrease) with
increasing $B$.
In Fig. \ref{circr}(d,e) we display the  probability density and probability density current
for the first positive-energy spin-down states of  $K$ and $K'$ valleys.
The results were obtained with the atomistic tight binding, in particular the current distribution was calculated using Eq. (4). The arrows representing the currents show the net currents
calculated by summation of interatomic currents on a square mesh of a side length of 2.7 nm. 
We find that the current orientation depends only on the valley, and that the current circulation 
in the $K'$ ($K$) valley leaves the negative (positive) potential on the left-hand side of the 
current orientation in agreement with the nature of the chiral confinement of the zero lines
in silicene \cite{szufran}. 

For quantitative analysis we define the current moment
\begin{equation} \chi=\frac{1}{2} \sum_{kl} \left.{\bf r}_{kl} \times {\bf J}_{kl}\right|_z,\label{chi}\end{equation}
where ${\bf r}_{kl}=({\bf r}_{k}+{\bf r}_{l})/2$ is the center of the bond between ion $k$ and ion $l$ and ${\bf J}_{kl}$ is the probability current flowing from ion $k$ to ion $l$
as given by Eq. (\ref{pro}). $\chi$ is negative (positive) for clockwise (counterclockwise) probability current flow. The magnetic dipole moment has the opposite orientation to $\chi$.

In Fig. 4(a) we plotted the values of $\chi$ for twenty energy levels of Fig. 3(a) of the lowest absolute value of the energy. We can see that the $\chi$ values are nearly the same for
all the states of fixed valley, i.e. the same current orientation and a very similar distribution of the currents is found for all the localized states of a given valley. 
In Fig. 4(b) and (c) we
displayed the zoom of the parts of Fig. 4(a) that correspond to $K$ and $K'$ valley states respectively. Splitting of the current moment with respect to the spin of the state can be resolved. 

 The change of the energy levels of Fig. 3(a,c) with $B$ is consistent with the classical formula 
for the interaction of the magnetic dipole moment generated by the current loop with the external magnetic field. The counterclockwise probability  current ($\chi>0$) in $K'$ states produces clockwise charge current, that generates the magnetic dipole moment oriented to the $-z$ direction, i.e. anti-parallel to the external magnetic field, hence the increase of the confined $K'$ energy levels with growing $B$.
The orientation of the dipole moment and the sign of the energy change is opposite for the $K$ valley.

The structure of energy levels and the angular momentum quantum numbers are presented in 
Fig. \ref{circr}(c) which contains a zoom of 
the continuum spectrum Fig. \ref{circr}(b) for low absolute value of the energy. 
We can see that all the energy levels of the degenerate quadruple have the same value of $|m|$.
For the first energy level at the positive energy side the angular momentum quantum number $m$ is equal to 0 for all the four states.
The $m$ values increase (decrease) by 1 for $K$ ($K'$) valley when one moves to quadruplets
of increasing energy.


\subsection{Circular potential: trivial confinement}

The properties of the states with topological confinement can be compared to the ones
found for the trivial one, which appears at a local reduction of the energy gap. 
The trivial confinement potential is taken as the absolute value of the one
given by Eq. (\ref{po2}) 
\begin{equation}
V^t_A({\bf r})=|V_g \left(1-2\exp(-r^4/R^4)\right)|, \label{potr}
\end{equation}
with the potential on the $B$ sublattice taken opposite
$V^t_B({\bf r})=-V^t_B({\bf r})$. Same $V_g$ and $R$ as in Eq. (1) are adopted.
The potential is plotted in Fig. \ref{trivial}(a).
The energy spectrum calculated with the tight binding approach is given Fig. \ref{trivial}(c).
For $|E|>V_g$ a large number of delocalized energy levels are observed. 
For $|E|<V_g$ a discrete energy level appear that are localized
near the dip of the local energy gap Fig. \ref{trivial}(a), with a clear separation 
of the conduction and valence bands in the spectrum. In Fig. \ref{trivial}(d) we plot
the results of the continuum approach with a zoom of the conduction band side of the spectrum
in Fig. \ref{trivial}(b). 

For larger $B$ the states near the energy gap [Fig. \ref{trivial}(d)] correspond to $K$ ($K'$) states at $E>0$ ($E<0$) side of the gap.
However, there is no general  strict correspondence between the valley index and the reaction of the
energy level to the change of the magnetic field which is observed for the topological confinement of the precedent subsection.
In Fig. \ref{trivial}(b) one can find the localized states which move up and down the energy scale for any valley. Moreover, for a given energy level the sign of $dE/dB$ derivative changes with $B$ -- see Fig. \ref{trivial}(c). The sign of this derivative agrees with the current moment as calculated with Eq. (\ref{chi}).

In Fig. \ref{trivpra} the current distribution is plotted for $K\downarrow$ $m=1$ state at $B=2$ T, $4$ T and $7.5$ T [see the dots in Fig. \ref{trivial}(b)]. 
In Fig. \ref{chi9} we plotted the values of the current moment $\chi$ for 30 states of the lowest
absolute values of the energy. The larger red, black, and purple dots show the values for the lowest-energy states at $E>0$ side. The lowest-ones correspond to the $K\downarrow$ states with the values of the $m$ quantum number given in Fig. \ref{chi9}. 
In Fig. \ref{trivpra} and in Fig. \ref{chi9} 
we can see that the orientation 
of the current for the $m=1$ $K\downarrow$ states is inverted between 2 T and  7.5 T, with the generated magnetic dipole moment reoriented from parallel to anti-parallel to the external magnetic field, respectively. 

For the topological confinement of the precedent section no reorientation of the current is observed with $B$. For the states confined at the flip of the electric field the current orientation is fixed by the valley degree of freedom,
 and the dependence of the confined energy levels on the magnetic field is monotonic.

In circular semiconductor quantum rings \cite{add1,add2} the ground-state of a single electron at $B=0$ 
corresponds to zero angular momentum  and is only degenerate with respect to the spin. For this state the persistent charge current at $B=0$ is zero \cite{add1,add2},
and a nonzero value would break the inversion symmetry of the system.  The persistent current for this ground  state appears only induced by external field \cite{add2}. 
For the rings considered here as well as for graphene quantum rings without the Rashba interaction 
\cite{trin3} the lowest-positive-energy states at $B=0$ are two-fold degenerate with
respect valley. Each of the valley degenerate states corresponds to nonzero but opposite persistent current and the magnetic field lifts the degeneracy of the states due to opposite sign of the magnetic dipole moments for these states.  
According to Ref. \cite{berry} the valley crossings in the magnetic field which are well visible in the spectra for the topological confinement, correspond to magnetic fields for which the Berry phase is an integer multiple of $\pi$.

The states studied here for both the trivial and the topological confinement are bound 
and are similar in this respect to the quantum dots for which the energy levels can be resolved
by the transport spectroscopy \cite{eqd}. The persistent currents can be deduced from the dependence
of the energy levels on the external perpendicular magnetic field.

\subsection{Coupled current loops}
The results of this section are obtained for topological confinement with twin centers given by Eq. (\ref{ala}) (see Fig. \ref{siatka}).
The calculated energy spectra are displayed in Fig. \ref{xplain} for $l=R$ (a,b) and $l=1.2R$  (c,d).
For $l=R$ the zero line forms a single loop [see Fig. \ref{siatka}]. The probability current for the lowest positive energy  $K'\downarrow$ level at $B=0$ is plotted in Fig. \ref{next}(a). The energy spectrum [Fig. \ref{xplain}(a,b)] resembles the one of the
single ring [Fig. \ref{circr}(a,b)] only with a larger number of bound energy levels.

For the distance between the ring centers increased to $l=1.2R$ the energy levels 
at $B=0$ become nearly two-fold degenerate for each spin and valley [Fig. \ref{circr}(c,d)]. 
The rings are nearly separated and
a tunnel coupling between them is present [Fig. \ref{next}(b)].
The electron density bears signatures of  bonding [Fig. \ref{next}(b)] and antibonding [Fig. \ref{next}(c)] orbitals with enhanced and reduced tunneling between  separate rings.  Figure \ref{circr}(d) shows that at higher $B$ the $K'$ states at $E>0$ and $K$ energy levels at $E<0$ 
tend to degenerate in pairs, which indicates lifting of the tunnel coupling between the 
rings. In order to study this effect in more detail in Fig. \ref{fju}(a) and (b) we plotted the spin-down $K$ and $K'$ energy levels, respectively,
with the color of the lines that indicates the average distance to the nearest 
center of the ring $\langle r'\rangle$, where for position ${\bf r}$ in space
$r'$ is defined as $r'=\min(|{\bf r}-(l,0,0)|,|{\bf r}+(l,0,0)|)$.
We can see that for $B=0$ the localization of the wave functions 
measured with $r'$ is the strongest for the states near the zero energy.
As the magnetic field grows, the average $r'$ is decreased for all the states.
For high magnetic field  ($B=10$ T) the strongest localization is observed
for the localized $K$  levels of the lowest energy and for the $K'$ energy levels of the highest energy. When $r'$ is small the densities are more strongly localized near the centers of separate rings so the tunnel coupling between the rings is lifted and the energy levels 
become double degenerate. 

\section{Summary}
We have studied the states bound by inhomogenous vertical electric field in buckled silicene 
that is either reduced or inverted along a closed line that supports trivial and topological carrier confinement, respectively.
We used the atomistic tight-binding approach and the continuum model for both radially
symmetric systems and for pairs of coupled inversion loops. We determined the discrete
part of the spectrum within the energy gap that is open by the vertical electric field far from its inversion area.
For trivial confinement the orientation of the persistent currents depends on the external magnetic field and can be counterclockwise or clockwise for both valleys. For the topological confinement the orientation of the persistent current is fixed by the valley degree of freedom.
 
\section*{Acknowledgments}
This work was supported by the National Science Centre (NCN) according to decision DEC-2016/23/B/ST3/00821.The calculations were performed
on PL-Grid Infrastructure and the Prometheus server at ACK Cyfronet AGH.

\end{document}